\documentclass[tightenlines,preprint,aps,prd,showpacs,nofootinbib]{revtex4}
\usepackage{graphicx}
\usepackage{amsmath}
\usepackage{amssymb}
\usepackage{bm}

\begin{document}

%%%%%%%%%%%%%%%%%%%%%%%%%%%%%%%%%%%%%%%%%%%%%%%%%%%%%%%%%%%%%%%%%%%%%%%%%%%%%%
%Title of paper
\title{\mbox{}\\[10pt]
Factorization in the Production and Decay of the $X(3872)$ 
}
%%%%%%%%%%%%%%%%%%%%%%%%%%%%%%%%%%%%%%%%%%%%%%%%%%%%%%%%%%%%%%%%%%%%%%%%%%%%%%

\author{Eric Braaten and Masaoki Kusunoki}
%\email[]{Your e-mail address}
%\homepage[]{Your web page}
%\thanks{}
%\altaffiliation{}
\affiliation{
Physics Department, Ohio State University, 
Columbus, Ohio 43210, USA}

\date{\today}
%%%%%%%%%%%%%%%%%%%%%%%%%%%%%%%%%%%%%%%%%%%%%%%%%%%%%%%%%%%%%%%%%%%%%%%%%%%%%%
\begin{abstract}
% insert abstract here
The production and decay of the $X(3872)$ are analyzed under the
assumption that the $X$ is a weakly-bound molecule 
of the charm mesons $D^0 \bar D^{*0}$ and $D^{*0} \bar D^0$.
The decays imply that the large $D^0 \bar D^{*0}$ scattering length
has an imaginary part.
An effective field theory for particles 
with a large complex scattering length is used to derive factorization 
formulas for production rates and decay rates of $X$.
If a partial width is calculated in a model with a particular
value of the binding energy, the factorization formula can be used 
to extrapolate to other values of the binding energy and to take 
into account the width of the $X$.
The factorization formulas relate the rates for
production of $X$ to those for production of 
$D^0 \bar D^{*0}$ and $D^{*0} \bar D^0$ near threshold.  
They also imply that the line shape of $X$ 
differs significantly from that of a Breit-Wigner resonance.
\end{abstract}

%%%%%%%%%%%%%%%%%%%%%%%%%%%%%%%%%%%%%%%%%%%%%%%%%%%%%%%%%%%%%%%%%%%%%%%%%%%%%%
% insert suggested PACS numbers in braces on next line
% 12.38.-t   Quantum chromodynamics
% 12.39.St  Factorization
% 13.20.Gd  Decays of J/psi, Upsilon, and other quarkonia
% 14.40.Gx  Mesons with S=C=B=0, mass > 2.5 GeV (including quarkonia)

%%%%%%%%%%%%%%%%%%%%%%%%%%%%%%%%%%%%%%%%%%%%%%%%%%%%%%%%%%%%%%%%%%%%%%%%%%%%%%
% insert suggested keywords - APS authors don't need to do this
%\keywords{}

%%%%%%%%%%%%%%%%%%%%%%%%%%%%%%%%%%%%%%%%%%%%%%%%%%%%%%%%%%%%%%%%%%%%%%%%%%%%%%
%\maketitle must follow title, authors, abstract, \pacs, and \keywords
\maketitle

%%%%%%%%%%%%%%%%%%%%%%%%%%%%%%%%%%%%%%%%%%%%%%%%%%%%%%%%%%%%%%%%%%%%%%%%%%%%%%
% body of paper here - Use proper section commands
% References should be done using the \cite, \ref, and \label commands

%%
\section{Introduction}

The $X(3872)$ is a narrow resonance near $3872$ MeV discovered 
by the Belle collaboration in 2003 \cite{Choi:2003ue}. 
It has been observed through the exclusive decay 
$B^\pm\to XK^\pm$  \cite{Choi:2003ue,Aubert:2004ns} and
through its inclusive production in proton-antiproton 
collisions \cite{Acosta:2003zx,Abazov:2004kp}.
Its mass $m_X$ is extremely close to the threshold for the 
charm mesons $D^{0}$  and $\bar D^{*0}$ \cite{Olsen:2004fp}: 
%-----------------
\begin{equation}
m_{X} - (m_{D^{0}}+m_{D^{*0}})= +0.6\pm 1.1  \; {\rm MeV}.
\label{mXdiff}
\end{equation}
%-----------------
The upper bound on its 
decay width $\Gamma_X$ is \cite{Choi:2003ue}  
%-----------------
\begin{eqnarray}
 \Gamma_X < 2.3 \;{\rm MeV} \;\;\; {\rm ( 90 \% \; C.L. )}.
 \label{widthX} 
\end{eqnarray}
%-----------------
The $X(3872)$ was discovered through its decay into
$J/\psi \, \pi^+\pi^-$ \cite{Choi:2003ue}.
The Belle collaboration has recently presented evidence for the
decays $X \rightarrow J/\psi \, \pi^+\pi^-\pi^0$ 
and $X\to J/\psi \, \gamma$ \cite{Abe:2005ix}.
The decay $X\to J/\psi \, \gamma$ implies that the $X$ 
has positive charge conjugation.
Upper bounds have been set on several other decay modes 
\cite{Abe:2003zv,Yuan:2003yz,Aubert:2004fc,Abe:2004sd,Metreveli:2004px}.  

The nature of the $X(3872)$ has not yet been definitively established.
The presence of the $J/\psi$ among its decay products 
motivates its interpretation as a charmonium state with constituents 
$c\bar c$ \cite{Barnes:2003vb,Eichten:2004uh,Quigg:2004nv}.
Two interpretations that are motivated by the proximity of $m_X$ 
to the $D^{0}\bar D^{*0}$ threshold 
are a hadronic molecule with constituents $D D^*$
\cite{Tornqvist:2003na,Tornqvist:2004qy,Voloshin:2003nt,Wong:2003xk,%
Braaten:2003he,Swanson:2003tb,Swanson:2004pp} 
and a ``cusp'' associated with the $D^{0}\bar D^{*0}$ threshold 
\cite{Bugg:2004rk,Bugg:2004sh}.
Other proposed interpretations include 
a tetraquark with constituents $c \bar c q \bar q$ \cite{Vijande:2004vt},
a ``hybrid charmonium'' state with constituents $c \bar c g$
\cite{Close:2003mb,Li:2004st}, 
a glueball with constitutents $ggg$ \cite{Seth:2004zb},
and a diquark-antidiquark bound state with constituents $c u + \bar c \bar u$ 
\cite{Maiani:2004vq}.
Measurements of the decays of the $X$ can be used to determine 
its quantum numbers and narrow down the options 
\cite{Close:2003sg,Pakvasa:2003ea,Rosner:2004ac,Kim:2004cz,Abe:2005iy}.
The most predictive of the proposed interpretations are 
charmonium and $D D^*$ molecules.
The $C=-$ charmonium options are ruled out by the decay 
$X\to J/\psi \, \gamma$.  Evidence ruling out or disfavoring 
each of the $C=+$ charmonium options
has been accumulating \cite{Olsen:2004fp, Quigg:2004vf,Abe:2005iy}.
The most difficult charmonium state to rule out is the $\chi_{c1}(2P)$,
partly because it can have resonant S-wave interactions with 
$D^{0}\bar D^{*0}$ and $D^{*0}\bar D^{0}$ that transform it into 
a $D D^*$ molecule \cite{Braaten:2003he}. 

The possibility that charm mesons might form molecular states
was considered shortly after the discovery of charm 
\cite{Bander:1975fb,Voloshin:ap,DeRujula:1976qd,Nussinov:1976fg}.  
The first quantitative study of the possibility of
molecular states of charm mesons was carried out by Tornqvist 
in 1993 using a one-pion-exchange potential model.
He found that the isospin-0 combinations of $D \bar D^*$ and $D^* \bar D$ 
could form weakly-bound states in the S-wave $1^{++}$ channel 
and in the P-wave $0^{-+}$ channel \cite{Tornqvist:1993ng}. 
Since the binding energy is small compared to the 8.4 MeV splitting 
between the $D^{0}  \bar D^{*0}$ threshold and the  $D^+ D^{*-}$ 
threshold, there are large isospin breaking effects 
\cite{Tornqvist:2003na,Tornqvist:2004qy}.
After the discovery of the $X(3872)$, Swanson considered a potential 
model that includes both one-pion-exchange and quark exchange,
and found that the $C=+$ superposition of $D^{0}  \bar D^{*0}$
and $D^{*0} \bar D^{0}$ could form a weakly-bound state 
in the S-wave $1^{++}$ channel \cite{Swanson:2003tb}.
Another mechanism for generating a $D  D^*$ molecule is the 
accidental fine-tuning of the mass of the $\chi_{c1}(2P)$ to the 
$D^{0}  \bar D^{*0}$/$D^{*0} \bar D^{0}$ threshold which creates a 
$D  D^*$ molecule with quantum numbers $1^{++}$ \cite{Braaten:2003he}. 

The assumption that the $X(3872)$ is a weakly-bound 
$DD^*$ molecule is very predictive \cite{Braaten:2003he}.
This assumption has been used by Voloshin to predict the rates
and momentum distributions for the decays of $X$ into 
$D^0 \bar D^0 \pi^0$ and $D^0 \bar D^0 \gamma$ \cite{Voloshin:2003nt}.
It has been used to calculate the rate for the exclusive decay 
of $\Upsilon(4S)$ into the $X$ and two light hadrons \cite{Braaten:2004rw},
to estimate the decay rate for the discovery mode 
$B^+ \to X K^+$ \cite{Braaten:2004fk}, 
and to predict the suppression of the decay rate for 
$B^0 \to X K^0$ \cite{Braaten:2004ai}.
The assumption that the $X$ is a weakly-bound $D  D^*$ molecule
also has implications for inclusive production
\cite{Braaten:2004jg,Voloshin:2004mh}.

In this paper, we point out that if $X$ is a loosely-bound molecule,
its short-distance decay rates and its exclusive production rates
satisfy simple factorization formulas.
In Section~\ref{sec:2chmodel}, we illustrate the factorization formulas
using a two-channel scattering model.
In Section~\ref{sec:Xdecay}, we show how the rate
for a short-distance decay mode of the $X$, such as 
$J/\psi \, \pi^+\pi^-\pi^0$, can be factorized into 
long-distance factor that involves the large scattering length 
and a short-distance factor that is insenstitive to $a$.
In Section~\ref{sec:BtoXK}, we show how a production rate for the $X$,
such as the decay rate for $B^+ \to X K^+$,
can be factorized into a long-distance factor and a short-distance factor.
In Section~\ref{sec:BDDstar}, we use factorization to calculate the shapes of
the invariant mass distributions of $D^{0}\bar D^{*0}$ and  
$D^{*0} \bar D^0$ near threshold.
In Section~\ref{sec:Xlineshape}, we use factorization to calculate 
the line shape of the $X$ in any of its short-distance decay modes.  
The line shape depends on the real and 
imaginary parts of the scattering length and can differ
substantially from a conventional Breit-Wigner resonance.
A summary of our results is given in Section~\ref{sec:summary}.

\section{Factorization in a Simple Scattering Model} 
\label{sec:2chmodel}

\subsection{Universality for large scattering length}

Nonrelativistic few-body systems with short-range interactions
and a large scattering length $a$ have universal properties that
depend on the scattering length but are otherwise insensitive  
to details at distances small compared to $a$ \cite{Braaten:2004rn}. 
In any specific system, there is a natural momentum scale $\Lambda$ 
that sets the scale of most low-energy scattering parameters.
The scattering length is large if it satisfies $|a| \gg \Lambda^{-1}$.
Universality predicts that the T-matrix element
for 2-body elastic scattering with relative momentum $p \ll \Lambda$ is
%-----------------
\begin{equation}
{\cal T}(p) = {2 \pi/\mu \over - 1/a - i p },
\label{T-uni}
\end{equation}
%-----------------
where $\mu$ is the reduced mass of the two particles.
If $a$ is real and positive, 
universality predicts that there is a weakly-bound state with
binding energy  
%-----------------
\begin{equation}
E_X= \frac{1}{2\mu a^2}.
\label{Eb}
\end{equation}
%-----------------
The universal momentum-space wavefunction of this bound state is
%-----------------
\begin{eqnarray}
\psi(p) = \frac{(8\pi/a)^{1/2}}{p^2+1/a^2}.
\label{psi-uni}
\end{eqnarray}
%-----------------
The universal amplitude for transitions from the bound state
to a scattering state consisting of two particles with relative 
momentum $p \ll \Lambda$ is
%-----------------
\begin{equation}
{\cal A}_X = {\sqrt{2 \pi} \over \mu} \, a^{-1/2}.
\label{AX-uni}
\end{equation}
%-----------------
These results are all encoded in the universal expression 
for the truncated connected transition amplitude:
%-----------------
\begin{equation}
{\cal A}(E) = {2 \pi/\mu \over - 1/a + \sqrt{-2 \mu E} }.
\label{A-uni}
\end{equation}
%-----------------

%------------------------------------------------------------------------------------
\begin{figure}[t]
\includegraphics[width=8cm]{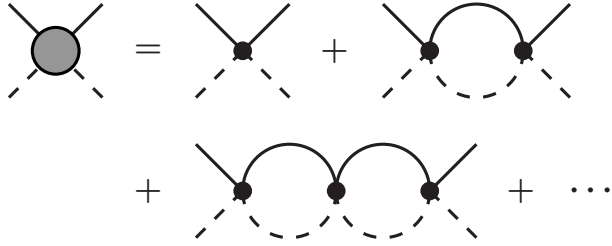}
\caption{
The geometric series of Feynman diagrams whose sum is the universal 
amplitude ${\cal A}(E)$. 
\label{fig:A-series}}
\end{figure}
%------------------------------------------------------------------------------------

The universal amplitude in Eq.~(\ref{A-uni}) can be obtained 
from a local effective field theory for the two particles.
The particles interact through an S-wave contact interaction 
with Feynman rule $- i (2 \pi/\mu) a_0(\Lambda)$, where the parameter 
$a_0(\Lambda)$ is a bare scattering length 
that depends on the ultraviolet momentum cutoff $\Lambda$.
The amplitude ${\cal A}(E)$ can be obtained 
by summing the geometric series of Feynman diagrams in 
Fig.~\ref{fig:A-series}:
%-----------------
\begin{equation}
{\cal A}(E) = -  
{(2 \pi/\mu) a_0(\Lambda) \over 1 + (2 \pi/\mu) a_0(\Lambda) L(\Lambda,E) },
\label{A-bare}
\end{equation}
%-----------------
where $L(\Lambda,E)$ is the amplitude for the propagation 
of the particles between successive contact interactions:
%-----------------
\begin{equation}
L(\Lambda,E) = {\mu \over \pi^2} 
\left( \Lambda - {\pi \over 2} \sqrt{- 2 \mu E} \right) .
\label{L-loop}
\end{equation}
%-----------------
Renormalization is accomplished by eliminating $a_0(\Lambda)$ 
in favor of the physical scattering length:
%-----------------
\begin{eqnarray}
a = {a_0(\Lambda) \over 1 + (2/\pi) \Lambda a_0(\Lambda) } .
\label{a-a0}
\end{eqnarray}
%-----------------
With this substitution, the expression for ${\cal A}(E)$ 
in Eq.~(\ref{A-bare}) reduces without approximation 
to the universal result in Eq.~(\ref{A-uni}).
Note that the scattering length $a$ can be tuned to $\pm \infty$
by tuning the bare scattering length to a critical value
of order $\Lambda^{-1}$:
%-----------------
\begin{eqnarray}
a_0(\Lambda) \longrightarrow - {\pi \over 2} \Lambda^{-1} .
\label{a0-tune}
\end{eqnarray}
%-----------------

If the 2-body system has inelastic scattering channels,
the large scattering length $a$ will have a negative imaginary part.
It is convenient to express the complex scattering length
in the form
%-----------------
\begin{eqnarray}
 \frac{1}{a}= \gamma_{\rm re} + i \gamma_{\rm im},
\label{a-complex}
\end{eqnarray} 
%-----------------
where $\gamma_{\rm re}$ and $\gamma_{\rm im}$ are real 
and $\gamma_{\rm im}\ge 0$.  In the case $\gamma_{\rm re}>0$
where there is a weakly-bound state, it can decay into the inelastic channel.
The expression for the binding energy on the right side of 
Eq.~(\ref{Eb}) is complex-valued.  Its real part $E_{X,{\rm pole}}$
and its imaginary part $\Gamma_X/2$ are given by
%-----------------
\begin{subequations}
\begin{eqnarray}
E_{X,{\rm pole}} &=& 
( \gamma_{\rm re}^2 - \gamma_{\rm im}^2 )/(2\mu),
\\
\Gamma_X  &=& 
2 \gamma_{\rm re} \gamma_{\rm im}/\mu .
\label{Gam-uni}
\end{eqnarray}
\end{subequations}
%-----------------
These quantities specify that there is a pole in the S-matrix 
at the energy $E = - E_{X,{\rm pole}} - i \Gamma_X/2$.
As we shall see in Section~\ref{sec:Xlineshape}, $\Gamma_X$ can be 
interpreted as the full width at half-maximum of
a resonance in the inelastic channel provided 
$\gamma_{\rm im} < \gamma_{\rm re}$.
The peak of the resonance is below 
the threshold for the two particles by 
%-----------------
\begin{eqnarray}
E_X &=& 
\gamma_{\rm re}^2/(2\mu) .
\label{EX-complex}
\end{eqnarray}
%-----------------
We therefore interpret $E_X$ as the binding energy
rather than $E_{X,{\rm pole}}$.

\subsection{Two-channel model}

Cohen, Gelman, and van Kolck have constructed a renormalizable 
effective field theory that describes two scattering channels 
with S-wave contact interactions \cite{Cohen:2004kf}.
We will refer to this model as the {\it two-channel scattering model}.
An essentially equivalent model has been used to describe the effects 
of $\Delta \Delta$ states on the two-nucleon system \cite{Savage:1996tb}.
The parameters of this model can be tuned to produce a large 
scattering length in the lower energy channel.  It can therefore be used as
a simple model for the effects on the $D^0 \bar D^{*0}$/$D^{*0} \bar D^0$
system of other hadronic channels with nearby thresholds, 
such as $J/\psi \, \rho$, $J/\psi \, \omega$, and $D^\pm D^{*\mp}$.

The two-channel model of Ref.~\cite{Cohen:2004kf} 
describes two scattering 
channels with S-wave contact interactions only.
We label the particles in the first channel $1a$ and $1b$
and those in the second channel $2a$ and $2b$.
We denote the reduced masses in the two channels by $\mu_1$ and $\mu_2$.
Renormalized observables in the 2-body sector are expressed in terms of 
4 parameters: three interaction parameters $a_{11}$, $a_{12}$, 
and $a_{22}$ with dimensions of length and the
energy gap $\Delta$ between the two scattering channels,
which is determined by the masses of the particles:
%-----------------
\begin{eqnarray}
\Delta &=& m_{2a} + m_{2b}- (m_{1a}+m_{1b}).
\label{Delta-real}
\end{eqnarray} 
%-----------------
The scattering parameters in Ref.~\cite{Cohen:2004kf} were defined
in such a way that $a_{11}$ and $a_{22}$ reduce in the limit 
$a_{12} \to \pm \infty$ to the scattering lengths for the two channels.  
The truncated connected transition amplitude 
${\cal A}(E)$ for this coupled-channel system is 
a $2\times 2$ matrix that depends on the energy $E$ 
in the center-of-mass frame. 
If that energy is measured relative to the threshold $m_{1a}+m_{1b}$ 
for the first scattering channel, 
the inverse of the matrix ${\cal A}(E)$ is%
\footnote{The expression for the matrix $T_s^{-1}$ in Eq.~(2.18) 
of Ref.~\cite{Cohen:2004kf} should be equal to ${\cal A}(E)^{-1}$
evaluated at $E = p^2/(2 \mu_1)$.  There is an error in the
22 component of $T_s^{-1}$: the square root $\sqrt{p^2- 2 \mu_2 \Delta}$
should be $\sqrt{(\mu_2/\mu_1)p^2 - 2 \mu_2 \Delta}$.}
%-----------------
\begin{eqnarray}
{\cal A}(E)^{-1}= \frac{1}{2\pi}
\begin{pmatrix}
\mu_1 \big[ -1/a_{11} +\sqrt{- 2 \mu_1 E} \, \big]  & 
	\sqrt{\mu_1 \mu_2}/a_{12}   \\
\sqrt{\mu_1 \mu_2}/a_{12} & 
	\mu_2 \big[ -1/a_{22}  + \sqrt{2 \mu_2 (\Delta - E)} \, \big]
\end{pmatrix}. 
\label{A-2cm}
\end{eqnarray}
%-----------------
The square roots are defined for negative real arguments by the 
prescription $E \to E + i \epsilon$ with $\epsilon \to 0^+$.
The amplitudes defined by Eq.~(\ref{A-2cm}) are for transitions between states 
with the standard nonrelativistic normalizations.  The transitions 
between states with the standard relativistic normalizations
are obtained by multiplying by a factor $\sqrt{2 m_i}$ for every 
particle in the initial and final state.
We will need explicit expression for the $11$ and $12$ entries 
of this matrix:
\begin{subequations}
\begin{eqnarray}
{\cal A}_{11}(E) &=& {2 \pi \over \mu_1} 
\left( - {1 \over a_{11}} + \sqrt{-2\mu_1 E} 
- {1 \over a_{12}^2} 
\Big[ -1/a_{22} + \sqrt{2 \mu_2 (\Delta -  E )} \, \Big]^{-1}  
\right)^{-1},
\label{A11-2cm}
\\
{\cal A}_{12}(E) &=& {2 \pi \over \sqrt{\mu_1 \mu_2}} 
\left( {1 \over a_{12}}
- a_{12} \left[ - {1 \over a_{11}} + \sqrt{-2\mu_1 E } \right]  
	\left[ - {1 \over a_{22}} + \sqrt{2\mu_2 (\Delta - E )} \right] 
\right)^{-1}. 
\label{A12-2cm}
\end{eqnarray}
\end{subequations}

The T-matrix element for the elastic scattering of particles 
in the first channel with relative momentum $p$ is obtained by evaluating
${\cal A}_{11}(E)$ in Eq.~(\ref{A11-2cm}) at the energy $E = p^2/(2 \mu_1)$:
%-----------------
\begin{eqnarray}
{\cal T}_{11}(p) = {2 \pi \over \mu_1} 
\left( - {1 \over a_{11}} - i p 
- {1 \over a_{12}^2} 
\Big[ -1/a_{22} + \sqrt{2 \mu_2 \Delta - (\mu_2/\mu_1)p^2} \, \Big]^{-1}  
\right)^{-1}.
\label{T11-2cm}
\end{eqnarray}
%-----------------
Setting ${\cal T}_{11}(0) = - 2 \pi/(\mu_1 a)$, we can read off the 
inverse scattering length $1/a$:
%-----------------
\begin{eqnarray}
{1 \over a} &=& \frac{1}{a_{11}}
+ \frac{1}{a_{12}^2} \Big[ \sqrt{2\mu_2 \Delta} - 1/a_{22} \, \Big]^{-1} .
\label{a-2cm}
\end{eqnarray}
%-----------------
%

If there is a bound state with energy $-\kappa^2/(2\mu_1)$
below the scattering threshold for the first channel,
the matrix ${\cal A}(E)$ given by Eq.~(\ref{A-2cm}) 
has a pole at $E = -\kappa^2/(2\mu_1)$.
The binding momentum $\kappa$ satisfies
%-----------------
\begin{eqnarray}
\kappa = \frac{1}{a_{11}}
+ \frac{1}{a_{12}^2} 
\Big[ - 1/a_{22} +\sqrt{2\mu_2 \Delta + (\mu_2/\mu_1) \kappa^2} \, \Big]^{-1}.
\label{kappa-2cm}
\end{eqnarray}
%-----------------
The behavior of the matrix ${\cal A}(E)$ as the energy $E$
approaches the pole associated with the bound state is
%-----------------
\begin{eqnarray}
{\cal A}(E) \longrightarrow - {1 \over E + \kappa^2/(2\mu_1)} 
	\begin{pmatrix} {\cal A}_{X1}\\ {\cal A}_{X2} \end{pmatrix}
  \otimes \begin{pmatrix} {\cal A}_{X1} & {\cal A}_{X2} \end{pmatrix}.
\label{AE-pole}
\end{eqnarray}
%----------------- 
The components ${\cal A}_{X1}$ and ${\cal A}_{X2}$ of the column vector are 
the amplitudes for transitions from the bound state to particles 
in the first and second channels, respectively.
The column vector is an eigenvector of the matrix ${\cal A}(E)^{-1}$ 
in Eq.~(\ref{A-2cm}) with eigenvalue zero, so its components must satisfy
%-----------------
\begin{eqnarray}
\mu_1 [-1/a_{11} + \kappa ] \, {\cal A}_{X1}
+ [\sqrt{\mu_1 \mu_2}/a_{12}] \, {\cal A}_{X2} = 0.
\end{eqnarray}
%-----------------

\subsection{Two-channel model with large scattering length}

The two-channel model of Ref.~\cite{Cohen:2004kf} can be used as a 
phenomenological model for a system with a large scattering length $a$
in the first channel.  The large scattering length requires a 
fine-tuning of the parameters $a_{11}$, $a_{22}$, $a_{12}$, and $\Delta$.
There are various ways to tune the parameters so that $a \to \pm \infty$.
For example, if $a_{11} < a_{12}^2/a_{22}$, the energy gap $\Delta$
can be tuned to the critical value where the right side of Eq.~(\ref{a-2cm})
vanishes.  Alternatively, the 
scattering parameter $a_{11}$ can be tuned to the critical value
$- a_{12}^2 [ \sqrt{2\mu_2 \Delta} - 1/a_{22} ]$.
The coefficients in the low-energy expansion of 
${\cal T}_{11}(p)^{-1}$ are proportional to various powers of 
$1/a_{11}$, $1/a_{12}$, and $\sqrt{2 \mu_2 \Delta}$.
We assume that these momentum scales are comparable in magnitude.
We refer to that common momentum scale as the natural low-energy scale 
and we denote it by $\Lambda$.

For $|a| \gg \Lambda^{-1}$ and $|E| \ll \Lambda^2/(2 \mu_1)$,
the amplitude ${\cal A}_{11}(E)$ in Eq.~(\ref{A11-2cm}) 
approaches the universal amplitude ${\cal A}(E)$
in Eq.~(\ref{A-uni}) with $\mu = \mu_1$.
It follows that for $p \ll \Lambda$ the T-matrix element 
${\cal T}_{11}(p)$ in Eq.~(\ref{T11-2cm}) approaches the
universal T-matrix element ${\cal T}(p)$ in Eq.~(\ref{T-uni}). 
For $a \gg \Lambda^{-1}$, the solution to Eq.~(\ref{kappa-2cm})
for the binding momentum $\kappa$ approaches $1/a$, so $\kappa^2/(2 \mu)$
approaches the universal binding energy $E_X$ in Eq.~(\ref{Eb}).
Finally the amplitude ${\cal A}_{X1}$ for transitions from the bound 
state to particles in the first channel, which is defined 
in Eq.~(\ref{AE-pole}), approaches the universal 
amplitude ${\cal A}_X$ in Eq.~(\ref{AX-uni}).

There are also universal features associated with transitions to the 
second channel.
If $|a| \gg \Lambda^{-1}$ and $|E| \ll \Lambda^2/\mu_1$, the leading term 
in the transition amplitude ${\cal A}_{12}(E)$ in Eq.~(\ref{A12-2cm})
reduces to 
\begin{eqnarray}
{\cal A}_{12}(E) = 
- {\sqrt{\mu_1/\mu_2}  \over a_{12}}
\Big[ \sqrt{2\mu_2 \Delta} - 1/a_{22} \, \Big]^{-1} 
{\cal A}(E),
\label{A12-largea}
\end{eqnarray}
where ${\cal A}(E)$ is the universal amplitude in 
Eq.~(\ref{A-uni}) with $\mu$ replaced by $\mu_1$.   
For $a \gg \Lambda^{-1}$, the leading term 
in the amplitude ${\cal A}_{X2}$ for transitions of the 
weakly-bound state $X$ to particles in the second channel,
which is defined in Eq.~(\ref{AE-pole}), reduces to
\begin{eqnarray}
{\cal A}_{X2} = 
- {\sqrt{\mu_1/\mu_2}  \over a_{12}}
\Big[ \sqrt{2\mu_2 \Delta} - 1/a_{22} \, \Big]^{-1} 
{\cal A}_X ,
\label{AX2-largea}
\end{eqnarray}
where ${\cal A}_X$ is the universal amplitude in Eq.~(\ref{AX-uni}). 
Note that the ratio ${\cal A}_{12}(E)/{\cal A}_{X2}$
of the amplitudes in Eqs.~(\ref{A12-largea}) and
(\ref{AX2-largea}) is a universal function of $a$ and $E$ only.

The expressions for ${\cal A}_{12}(E)$ and ${\cal A}_{X2}$ in 
Eqs.~(\ref{A12-largea}) and (\ref{AX2-largea}) are examples of 
{\it factorization formulas}.
They express the leading terms in the amplitudes as products 
of the same short-distance factor 
and different long-distance factors ${\cal A}(E)$ and ${\cal A}_X$.
The long-distance factors involve the large scattering length $a$.
The limit $|a| \to \infty$ has been taken in the short-distance factors.
The conditions $|E| \ll \Lambda^2/(2\mu)$ or $E=-E_X$
require the particles in the second channel to be off the 
energy shell by approximately $\Delta$.  In the short time 
$1/\Delta$ allowed by the uncertainty principle, those particles 
can propagate only over short distances of order  
$(2 \mu_2 \Delta)^{-1/2}$. This is small compared to the distance 
scales $(2 \mu |E|)^{-1/2}$ or $|a|$ associated with the particles 
in the first channel.  
Thus as far as they are concerned, the particles in the second 
channel act only as a point source for particles in the first channel.
The amplitudes for particles from such a point source to evolve into
particles of energy $E$ and into the weakly-bound state are 
$L(\Lambda,E) {\cal A}(E)$ and $L(\Lambda,-E_X) {\cal A}_X$,
respectively. By using the conditions $|E|,E_X \ll \Lambda^2/(2\mu)$, 
these amplitudes reduce to $(\mu \Lambda/\pi^2) {\cal A}(E)$ 
and $(\mu \Lambda/\pi^2) {\cal A}_X$, respectively.
In these expressions, the short-distance factors are identical
and the long-distance factors are the same as those in 
Eqs.~(\ref{A12-largea}) and (\ref{AX2-largea}).

\subsection{Unstable particle in the second channel}
\label{sec:unstable}

Now let us suppose one of the scattering particles in the second 
channel has a nonzero width.  We take that particle to be $2b$.
We assume that its width $\Gamma_{2b}$ arises from its decay into 
particles with relativistic momenta that are much greater than the 
ultraviolet cutoff $\Lambda_{\rm UV}$ that defines the domain 
of validity of the two-channel model.  The momenta of the decay 
products are therefore also much greater than $\sqrt{2 \mu_2 \Delta}$. 
We assume that $\Gamma_{2b}$ is small compared to the mass $m_{2b}$,
but not necessarily small compared to the energy gap $\Delta$
between the two channels.  This makes it necessary to take into 
account the contribution to the self-energy of particle $2b$ 
from the coupling to its decay products.

Taking into account the self-energy of particle $2b$ would modify the
term $\sqrt{2 \mu_2(\Delta-E)}$ in the inverse of the matrix of 
transition amplitudes given in Eq.~(\ref{A-2cm}). 
That term arises from the amplitude for the propagation of particles 
in the second channel between contact interactions, 
which is given by the integral
%-----------------
\begin{eqnarray}
\int {d^3p \over (2 \pi)^3} {-1 \over E-\Delta - p^2/(2 \mu_2) + i \epsilon}
= {\mu_2 \over \pi^2} \left( \Lambda_{\rm UV}
- {\pi \over 2} \sqrt{2\mu_2(\Delta - E - i \epsilon)} \right). 
\label{int-prop}
\end{eqnarray}
%-----------------
The cutoff constrains the momentum  to the region 
in which the nonrelativistic approximation for the energy of the 
particle $2b$ is valid.  In this region, the self-energy $\Pi$ can be
expressed as a function of $E'=E- \Delta - p^2/(2 \mu_2)$.
It can be taken into account by replacing $\Delta$ in the integral 
in Eq.~(\ref{int-prop}) by $\Delta + \Pi(E')$.
The assumption that the decay products of particle $2b$ have
relativistic momenta comparable to $m_{2b}$ implies that their 
contributions to $\Pi(E')$ have significant dependence on $E'$ 
only for variations in $E'$ that are comparable to $m_{2b}$.
For energies satisfying $|E| \ll \Lambda_{\rm UV}^2/(2 \mu_2)$ 
and loop momenta $p < \Lambda_{\rm UV}$, the dependence on $E'$ can be 
neglected and the argument of $\Pi(E')$ can be set to a constant, 
such as $- \Delta$.
The prescription for taking into account the self-energy then 
reduces to replacing $\Delta$ in the integral 
in Eq.~(\ref{int-prop}) by $\Delta + \Pi(-\Delta)$.
The real part of $\Pi(-\Delta)$ can be absorbed into $\Delta$ so that it
becomes the physical threshold.  The imaginary part of $\Pi(-\Delta)$ 
is related to the width of particle $2b$:  
${\rm Im} \Pi(-\Delta) = -  \Gamma_{2b}/2$.
Thus the  leading effect of the self-energy can be taken into
account by replacing $\Delta$ in Eq.~(\ref{A-2cm})
by the complex-valued energy gap
%-----------------
\begin{eqnarray}
\Delta &=& m_{2a} + m_{2b}- (m_{1a}+m_{1b}) - i \Gamma_{2b}/2 .
\label{Delta-complex}
\end{eqnarray} 
%-----------------

If $\Delta$ is complex, the solution to Eq.~(\ref{kappa-2cm}) 
for the binding momentum $\kappa$ is complex. 
It determines the pole mass $m_{X,{\rm pole}}$
and the width $\Gamma_X$ of the weakly-bound state according to
%----------------- 
\begin{eqnarray} 
m_{1a} + m_{1b} - \kappa^2/(2 \mu_1) = m_{X,{\rm pole}} - i \Gamma_X /2.
\label{mGam}
\end{eqnarray} 
%----------------- 
The imaginary part reflects the fact that the bound state 
can  decay into particle 2a and decay products of particle 2b.
The quantities $m_{X,{\rm pole}}$ and $\Gamma_X$ in Eq.~(\ref{mGam})
give the location of a pole in the S-matrix.  
They need not have the standard interpretations as the location 
of the peak and the full width 
at half maximum of a Breit-Wigner resonance.

\subsection{The $\bm{D^{0} {\bar D^{*0}}}$/$\bm{D^{*0} {\bar D^{0}}}$ System}
\label{sec:DDstar}

The energy difference between the mass of $X$ and the 
$D^{0} \bar D^{*0}$ threshold, which is given in Eq.~(\ref{mXdiff}),
is small compared to the natural energy scale for binding by
the pion exchange interaction: $m_{\pi}^2/2\mu \approx 10$ MeV, 
where $\mu$ is the reduced mass of $D^0$ and $\bar D^{*0}$.
The unnaturally small value of the energy difference 
implies that if the $X$ couples to 
$ D^{0} \bar D^{*0}$ and $D^{*0} \bar D^{0}$,
the S-wave scattering lengths for those channels
must be large compared to the natural length scale $1/m_{\pi}$
associated with the pion exchange interaction.
We assume that there is a large scattering length $a$ in the 
channel $(D D^*)^0_+$ with even charge conjugation defined by
%-----------------
\begin{eqnarray}
| (D D^*)^0_+ \rangle = {1 \over \sqrt{2}} 
\left( | D^0 \bar D^{*0} \rangle + | D^{*0} \bar D^0 \rangle \right) .
\end{eqnarray}
%-----------------
If the scattering length in the $C=-$ channel is negligible 
in comparison, the scattering lengths for elastic 
$D^{0} \bar D^{*0}$ scattering and elastic 
$D^{*0} \bar D^{0}$ scattering are both $a/2$.
We identify the $X$ as a bound state in the $(D D^*)^0_+$ channel.

The decays of the $X$ imply that the scattering length $a$ 
is complex-valued.  It can be parameterized in terms of the 
real and imaginary parts of $1/a$ as in Eq.~(\ref{a-complex}). 
Our interpretation of $X$ as a bound state requires $\gamma_{\rm re} > 0$.
The energy difference in Eq.~(\ref{mXdiff}) puts an upper bound on
$\gamma_{\rm re}$:
%-----------------
\begin{eqnarray}
\gamma_{\rm re} < 40 \ {\rm MeV} \hspace{1cm} ({\rm 90\% \ C.L.}).
\label{gamre:ub}
\end{eqnarray}
%-----------------
The upper bound on the width in Eq.~(\ref{widthX}) 
puts an upper bound on the product of $\gamma_{\rm re}$ and 
$\gamma_{\rm im}$:
%-----------------
\begin{eqnarray}
\gamma_{\rm re} \gamma_{\rm im} < (33 \ {\rm MeV})^2 
\hspace{1cm} ({\rm 90\% \ C.L.}).
\end{eqnarray}
%-----------------
There is also a lower bound on the width of the $X$
from its decays into $D^0 \bar D^0 \pi^0$ and $D^0 \bar D^0 \gamma$, 
which both proceed through the decay of a constituent $D^*$.
These decays involve interesting interference effects,
but the decay rates have smooth limits as the binding energy 
is tuned to 0 \cite{Voloshin:2003nt}.  In this limit, the constructive
interference increases the decay rate by a factor of 2:
the partial width of $X$ reduces to $2 \, \Gamma[D^{*0}]$.  
The width of $D^{*0}$ has not been measured, 
but it can be deduced from other information 
about the decays of $D^{*0}$ and $D^{*+}$.
Using the total width of the $D^{*+}$, 
its branching fraction into $D^+ \pi^0$, and isospin symmetry,
we can deduce the partial width of $D^{*0}$ into $D^0 \pi^0$:
$\Gamma[D^{*0} \to D^0 \pi^0] = 42 \pm 10$ keV.
The total width of the $D^{*0}$ can then be obtained by 
dividing by its branching fraction into $D^0 \pi^0$:
$\Gamma[D^{*0}] = 68 \pm 16$ keV.
The sum of the partial widths of $X$ into $D^0 \bar D^0 \pi^0$ and 
$D^0 \bar D^0 \gamma$ is therefore $136 \pm 32$ keV.
The resulting lower bound on the product 
of $\gamma_{\rm re}$ and $\gamma_{\rm im}$ is 
%-----------------
\begin{eqnarray}
\gamma_{\rm re} \gamma_{\rm im} > (7 \ {\rm MeV} )^2
\hspace{1cm} ({\rm 90\% \ C.L.}).
\end{eqnarray}
%-----------------
By combining this with the upper bound on $\gamma_{\rm re}$
in Eq.~(\ref{gamre:ub}), we can infer that $\gamma_{\rm im}> 1$ MeV.

\section{Short-distance decays of $X$} 
\label{sec:Xdecay}

The decay modes of the $X(3872)$ can be classified into 
{\it long-distance decays} and {\it short-distance decays}.
The long-distance decay modes are $D^0 \bar D^0 \pi^0$ and  
$D^0 \bar D^0 \gamma$, which proceed through the decay of a 
constituent $D^{*0}$ or $\bar D^{*0}$.  These decays are dominated 
by a component of the wavefunction of the $X$ in which the separation 
of the $D$ and $D^*$ is of order $1/|a|$.  
These long-distance decays involve interesting interference effects 
between the $D^0 \bar D^{*0}$ and $D^{*0} \bar D^0$ components 
of the wavefunction \cite{Voloshin:2003nt}.
The short-distance decays involve a component of the 
wavefunction in which the separation of the $D$ and $D^*$ 
is of order $1/m_\pi$ or smaller.  Examples are the observed decay modes
$J/\psi \, \pi^+ \pi^-$, $J/\psi \, \pi^+ \pi^- \pi^0$, 
and $J/\psi \, \gamma$.

Short-distance decays of the $X$ into a hadronic final state $H$
involve well-separated momentum scales.
The $D D^*$ wavefunction of the $X$ involves the momentum scale $1/|a|$  
set by the large scattering length.  The transition of the $D D^*$ 
to $H$ involves momentum scales $m_{\pi}$ or larger.
We will refer to momentum scales of order $1/|a|$ and smaller as 
{\it long-distance} scales and momentum scales of order $m_\pi$
and larger as {\it short-distance} scales.  We denote the arbitrary 
boundary between these two momentum regions by $\Lambda$.

The separation of scales $|a| \gg 1/m_\pi$ in the decay process 
$X \to H$ can be exploited through a factorization formula for the 
T-matrix element:
%-----------------
\begin{eqnarray}
{\cal T}[ X \to H] = 
\sqrt{2 m_X} \, {\cal A}_X \times {\cal A}_{\rm short} [(D D^*)^0_+ \to H] .
\label{XtoH:fact}
\end{eqnarray}
%-----------------
In the long-distance factor, ${\cal A}_X$ is the universal amplitude 
given in Eq.~(\ref{AX-uni}) and the factor of $\sqrt{2 m_X}$ 
takes into account the difference between the standard nonrelativistic
and relativistic normalizations of states.  If the complex scattering 
length is parameterized as in Eq.~(\ref{a-complex}), this factor is
%-----------------
\begin{eqnarray}
{\cal A}_X = {\sqrt{2 \pi} \over \mu} 
\left( \gamma_{\rm re} + i \gamma_{\rm im} \right)^{1/2}.
\label{AX-complex}
\end{eqnarray}
%-----------------
The short-distance factor ${\cal A}_{\rm short}$ in Eq.~(\ref{XtoH:fact}) 
is insensitive to $a$, and one can therefore take the
limit $|a| \to \infty$ in this factor. 
The factorization formula in Eq.~(\ref{XtoH:fact}) can serve as a
definition of the short-distance factor.  The content of the 
factorization statement then resides in the fact that,
up to corrections suppressed by powers of $1/(a m_\pi)$, the 
same short-distance factor appears in the factorization formula 
for the T-matrix element for the scattering process
$D^0 \bar D^{*0} \to H$  at energies $E$ near the 
$D^0 \bar D^{*0}$ threshold:
%-----------------
\begin{eqnarray}
{\cal T}[ D^0 \bar D^{*0} \to H] = {1 \over \sqrt{2}} \,
\sqrt{4 m_{D^0} m_{D^{*0}}} \, {\cal A}(E) \times 
{\cal A}_{\rm short} [(D D^*)^0_+ \to H] .
\label{DDtoH:fact}
\end{eqnarray}
%-----------------
In the long-distance factor, the $1/\sqrt{2}$ is the amplitude for  
$D^0 \bar D^{*0}$ to be in the channel $(D D^*)^0_+$ 
with the large scattering length, the factor of 
$\sqrt{4 m_{D^0} m_{D^{*0}}}$ takes into account the difference 
between the standard nonrelativistic and relativistic 
normalizations of states, and 
${\cal A}(E)$ is the universal amplitude 
given in Eq.~(\ref{A-uni}).  
If the complex scattering length is 
parameterized as in Eq.~(\ref{a-complex}), this factor is
%-----------------
\begin{eqnarray}
{\cal A}(E) = 
{2 \pi/\mu \over
- \gamma_{\rm re} - i (\gamma_{\rm im} + \sqrt{2 \mu E}) }.
\label{AE-complex}
\end{eqnarray}
%-----------------
The factorization formulas in Eqs.~(\ref{DDtoH:fact}) and (\ref{XtoH:fact})
are analogous to those in Eqs.~(\ref{A12-largea}) and
(\ref{AX2-largea}) for the two-channel model with a
large scattering length in the first channel.

The factorization formulas in Eqs.~(\ref{XtoH:fact}) and (\ref{DDtoH:fact})
can be motivated diagrammatically by separating virtual particles 
into {\it soft} particles and {\it hard} particles according to whether 
they are off their energy shells by less than or by more than 
$\Lambda^2/(2 \mu)$, where $\Lambda$ is the arbitrary momentum 
separating the long-distance scale $1/|a|$ and the short-distance 
scale $m_\pi$.  Any contribution from soft particles inside a 
subdiagram all of whose external legs are hard can be 
Taylor-expanded in the momentum of the soft particles, leading to 
suppression factors of $1/(a \Lambda)$.  The diagrams with the
fewest suppression factors will be ones that can be separated 
into a part for which all the internal lines are hard particles
and a part that involves only soft particles.   This separation
leads to the factorization formula.

%------------------------------------------------------------------------------------
\begin{figure}[t]
\includegraphics[width=8cm]{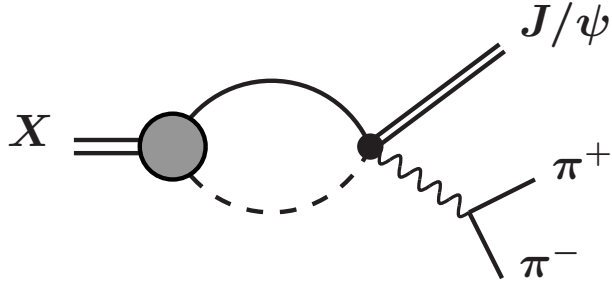}
\caption{
Feynman diagram for $X \to J/\psi \, \pi^+ \pi^-$
that scale like $(a m_\pi)^{-1/2}$.  
The $(DD^*)^0_+$ wavefunction of the $X$
is the integral over the loop energy of the product of the blob 
and the attached propagators.
\label{fig:XtoH}}
\end{figure}
%------------------------------------------------------------------------------------

The leading terms in the T-matrix element for $X \to H$
in the limit $|a| m_\pi \gg 1$ can be represented 
by the Feynman diagrams in Fig.~\ref{fig:XtoH} 
and can be expressed as
%-----------------
\begin{eqnarray}
{\cal T}[ X \to H] = \sqrt{2 m_X}
\int {d^3p \over (2 \pi)^3} \, \psi(p) \,
{\cal A}^{(\Lambda)} [(D D^*)^0_+ \to H] ,
\label{T-XH:Lam}
\end{eqnarray}
%-----------------
where $\psi(p)$ is the universal wavefunction in Eq.~(\ref{psi-uni}).
The factor ${\cal A}^{(\Lambda)}$, which is represented by a dot in 
Fig.~\ref{fig:XtoH}, is an amplitude for the transition $(D D^*)^0_+ \to H$ 
in which all virtual particles are
off their energy shells by more than $\Lambda^2/(2 \mu)$.
It is therefore insensitive to the relative momentum $\bm{p}$
of the $D$ and $D^*$.  If that momentum dependence is neglected
and if the integral in Eq.~(\ref{T-XH:Lam})
is regularized by a momentum cutoff $|\bm{p}| < \Lambda$,
the wavefunction factor reduces in the limit $|a| \gg \Lambda^{-1}$ to
\begin{eqnarray}
\int {d^3p \over (2 \pi)^3} \psi(p)
= \sqrt{2 \over \pi^3} \Lambda a^{-1/2}.
\label{psi-int}
\end{eqnarray}
The factorization formula in Eq.~(\ref{XtoH:fact}) is then obtained 
by absorbing a factor of $(\mu/\pi^2) \Lambda$ into ${\cal A}^{(\Lambda)}$
to obtain the short-distance factor:
\begin{eqnarray}
{\cal A}_{\rm short} [(D D^*)^0_+ \to H] =
\left( {\mu \over \pi^2} \Lambda \right) {\cal A}^{(\Lambda)} [(D D^*)^0_+ \to H].
\label{Ashort-DDtoH}
\end{eqnarray}
Since the T-matrix element in Eq.~(\ref{XtoH:fact}) is independent 
of the arbitrary separation scale, the dependence on $\Lambda$
must cancel between the two factors on the right side of 
Eq.~(\ref{Ashort-DDtoH}).

%------------------------------------------------------------------------------------
\begin{figure}[t]
\includegraphics[width=8cm]{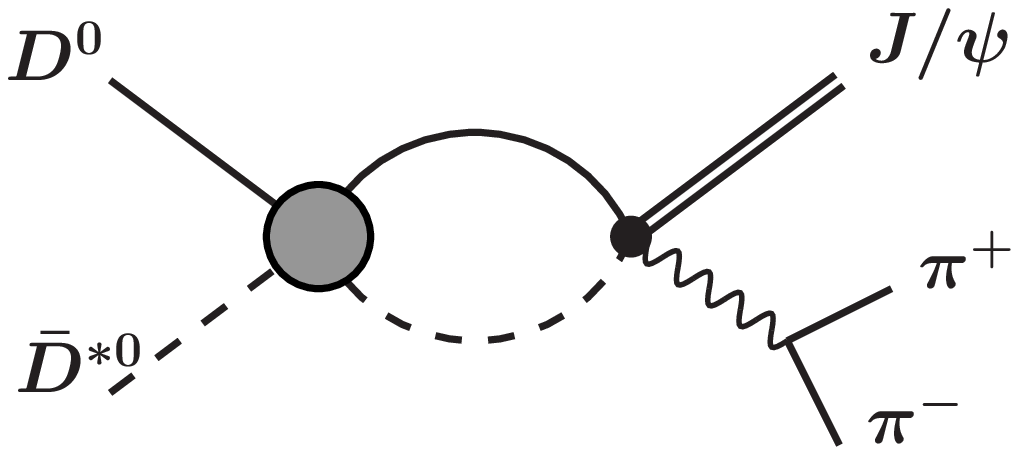}
\caption{
Feynman diagrams for $D^0 \bar D^{*0} \to J/\psi \, \pi^+ \pi^-$
that are enhanced near the $D^0 \bar D^{*0}$ threshold
by a factor of $a m_\pi$.  
The blob represents the geometric series of Feynman diagrams 
in Fig.~\ref{fig:A-series}.
\label{fig:DDtoH}}
\end{figure}
%------------------------------------------------------------------------------------

The leading term in the T-matrix element for $D^0 \bar D^{*0} \to H$
in the limits $|a| \gg \Lambda^{-1}$ and $E \ll \Lambda^2/(2\mu)$ 
can be represented by the Feynman diagram in 
Fig.~\ref{fig:DDtoH} and can be expressed as
\begin{eqnarray}
{\cal T}[ D^0 \bar D^{*0} \to H] = {1 \over \sqrt{2}} \, 
\sqrt{4 m_{D^0} m_{D^{*0}}} \, {\cal A}(E) \, L(\Lambda,E) \, 
{\cal A}^{(\Lambda)} [(D D^*)^0_+ \to H] .
\label{T-DDH:Lam}
\end{eqnarray}
The factor $L(\Lambda,E)$ is the amplitude for the
propagation of the $D$ and $D^*$ between successive contact 
interactions, which is given in Eq.~(\ref{L-loop}).
The approximation $E \ll \Lambda^2/(2\mu)$ justifies neglecting
the $\sqrt{2 \mu E}$ term in $L(\Lambda,E)$.  
The factorization formula in Eq.~(\ref{DDtoH:fact}) is then obtained 
by absorbing the remaining term $(\mu/\pi^2) \Lambda$ into 
${\cal A}^{(\Lambda)}$ to obtain the short-distance factor
in Eq.~(\ref{Ashort-DDtoH}).

The factorization formula for the T-matrix element in 
Eq.~(\ref{XtoH:fact}) implies a factorization formula for the 
decay rate for $X \to H$.  The decay rate $\Gamma[X \to H]$ 
can be expressed as the product of a
short-distance factor and the long-distance factor
%-----------------
\begin{eqnarray}
|{\cal A}_X|^2 = 
{2 \pi  \over \mu^2} \sqrt{\gamma_{\rm re}^2 + \gamma_{\rm im}^2}.
\label{AXsq}
\end{eqnarray}
%----------------- 
Using the expressions for the binding energy and the total width 
of the $X$ in Eqs.~(\ref{EX-complex}) and (\ref{Gam-uni}), 
the long-distance factor in Eq.~(\ref{AXsq}) can be expressed as
%-----------------
\begin{eqnarray}
|{\cal A}_X|^2 = 
\sqrt{8 \pi^2  \over \mu^3} [ E_X + \Gamma_X^2/(16 E_X) ]^{1/2}.
\label{AXsq-EGam}
\end{eqnarray}
%-----------------

If the partial width for a short-distance decay mode of the $X$ has been 
calculated using a model with a specific binding energy, 
the factorization formula for the decay rate can be 
used to extrapolate the prediction to other values of the binding energy
and to take into account the effect of the width of the $X$.
This is useful because numerical calculations in models often become 
increasingly unstable as the binding energy is tuned to zero.
Swanson has estimated the partial widths for various 
short-distance decays of $X$ using a potential model, but only for 
binding energies down to about 1 MeV and without taking into account 
the effect of the width of the $X$ \cite{Swanson:2003tb,Swanson:2004pp}.
His predictions can be extrapolated to other values of the binding energy 
and the width of the $X$ can be taken into account by using the 
long-distance factor in Eq.~(\ref{AXsq-EGam}).

\section{Production of $\bm{X}$}
\label{sec:BtoXK}

The production of $X$ necessarily involves the long-distance 
momentum scale $1/|a|$ through the $(DD^*)^0_+$ wavefunction of the $X$.
The production also involves much larger momentum scales.
Unless there are already hadrons in the initial state
containing a $c$ and $\bar c$, the production process involves 
the scale $m_c$ associated with the creation of a $c \bar c$ pair.
Even if the initial state includes hadrons that contain $c$ and $\bar c$,
such as $J/\psi$ or $D^+$ and $D^-$,
the production process involves the scale $m_\pi$ associated with the 
formation of the $D^0$ and $\bar D^{*0}$ that bind to form the  $X$.
We will define a {\it short-distance production} process to be
one for which the initial state either does not include any of the 
charm mesons $D^0$, $\bar D^{*0}$, $D^{*0}$, or $\bar D^0$, or if it does,
the momentum of the charm meson in the rest frame of the $X$ 
is of order $m_\pi$ or larger.
All practical production mechanisms for $X$ in high energy physics
are short-distance processes.  Long-distance production mechanisms 
could arise in a hadronic medium that includes charm mesons,
such as that produced by relativistic heavy-ion collisions.

In a short-distance production process,
the separation between the long-distance scale $1/|a|$ and all the 
shorter-distance momentum scales can be exploited through 
a factorization formula that expresses the leading term
in the production rate as the product of a long-distance factor 
that involves $a$ and a short-distance factor that is insensitive to $a$.
To be definite, we will consider the specific 
production process $B \to X K$.  The factorization for any other
short-distance production process will have the same long-distance factor
but a different short-distance factor.

There are many momentum scales that play an important role
in the decay $B \to X K$,  ranging from the extremely short-distance 
scales $m_W$ and $m_b$ associated with the quark decay process 
$b \to c \bar c s$ to the smaller short-distance scales 
$\Lambda_{\rm QCD}$ 
and $m_\pi$ involved in formation of the final-state hadrons
to the long-distance scale $1/|a|$ associated with the $(D D^*)^0_+$ 
wavefunction of the $X$.
We denote the arbitrary boundary between the long-distance momentum region
and the short-distance momentum region by $\Lambda$.

The separation between the long-distance scale $1/|a|$ and all the 
short-distance momentum scales in the decay $B \to X K$ can be exploited 
through a factorization formula for the T-matrix element:
%-----------------
\begin{eqnarray}
{\cal T}[B \to X K] =
{\cal A}_{\rm short}[B \to (DD^*)^0_+ K] \times {\cal A}_X \sqrt{2 m_X}.
\label{BtoXK:fact}
\end{eqnarray}
%-----------------
In the long-distance factor, ${\cal A}_X$ is the universal amplitude 
in Eq.~(\ref{AX-complex}).  The short-distance factor in 
Eq.~(\ref{BtoXK:fact}) is insensitive to $a$ and one can therefore take 
the limit $|a| \to \infty$ in that factor.
The factorization formula in (\ref{BtoXK:fact}) can serve as the 
definition of the short-distance factor.  The content of the 
factorization statement then resides in the fact that the 
same short-distance factor appears in the factorization formula 
for the T-matrix element for the decay $B \to D^0 \bar D^{*0} K$  
when the $D D^*$ invariant mass is near the 
$D^0 \bar D^{*0}$ threshold.  This factorization formula is 
discussed in Section~\ref{sec:BDDstar}.

%------------------------------------------------------------------------------------
\begin{figure}[t]
\includegraphics[width=8cm]{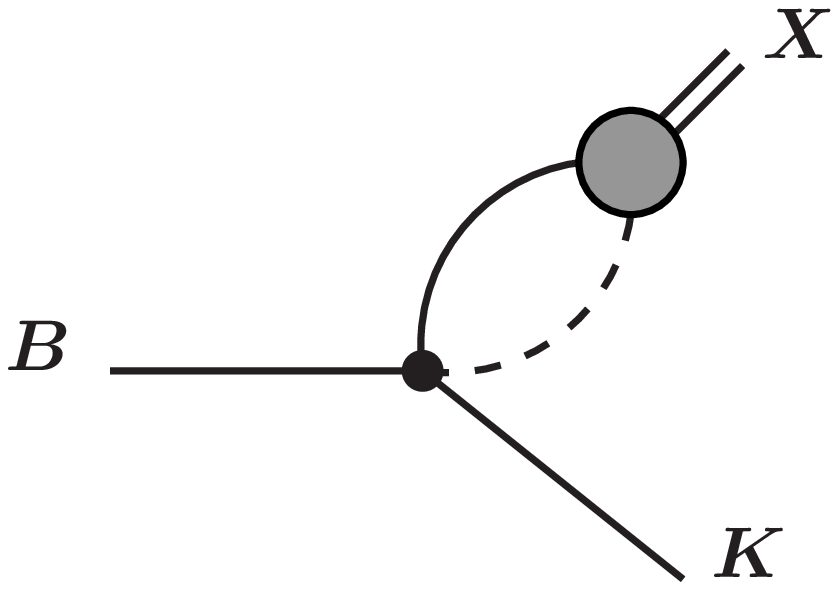}
\caption{
Feynman diagram for $B \to X K$ that scales like $(a m_\pi)^{-1/2}$.  
The $(DD^*)^0_+$ wavefunction of the $X$
is the integral over the loop energy of the product of the blob 
and the attached propagators.
\label{fig:BtoXK}}
\end{figure}
%------------------------------------------------------------------------------------

The factorization formula in Eq.~(\ref{BtoXK:fact}) can be motivated
diagrammatically by separating the loop integrals in the decay amplitude 
according to whether the  virtual particles are off their energy 
shells by less than or by more than $\Lambda^2/(2 \mu)$. 
The leading terms in the decay amplitude for $B \to X K$ 
are suppressed only by a factor of $(a m_\pi)^{-1/2}$.  
These terms can be represented by the Feynman diagram in 
Fig.~\ref{fig:BtoXK} and can be expressed in the form
%-----------------
\begin{eqnarray}
{\cal T}[B \to X K] =  \sqrt{2 m_X} \int {d^3p \over (2 \pi)^3} \psi(p) \, 
{\cal A}^{(\Lambda)}[B \to (DD^*)^0_+ K] ,
\label{T:BtoXK}
\end{eqnarray}
%-----------------
where $\psi(p)$ is the universal wavefunction in Eq.~(\ref{psi-uni}).
The factor ${\cal A}^{(\Lambda)}$, which is represented by a dot in 
Fig.~\ref{fig:BtoXK}, is an amplitude for the decay 
$B \to (DD^*)^0_+ K$ in which all virtual particles are
off their energy shells by more than $\Lambda^2/(2 \mu)$.
It is therefore insensitive to the relative momentum $\bm{p}$
of the $D$ and $D^*$.  If that momentum dependence is neglected,
the wavefunction factor in Eq.~(\ref{T:BtoXK}) reduces to 
Eq.~(\ref{psi-int}).
The factorization formula in Eq.~(\ref{BtoXK:fact}) then requires the
short-distance factor to be
%-----------------
\begin{eqnarray}
{\cal A}_{\rm short}[B \to (DD^*)^0_+ K] = 
{\cal A}^{(\Lambda)}[B \to (DD^*)^0_+ K]
\left( {\mu \over \pi^2} \Lambda \right).
\label{Ashort-BtoDDK}
\end{eqnarray}
%-----------------
Since the T-matrix element in Eq.~(\ref{BtoXK:fact}) is independent 
of the arbitrary separation scale, the dependence on $\Lambda$
must cancel between the two factors on the right side of 
Eq.~(\ref{Ashort-BtoDDK}).

We proceed to use the factored expression in Eq.~(\ref{BtoXK:fact})
to evaluate the decay rate for $B^+ \to X K^+$.
Lorentz invariance constrains the short-distance amplitude 
${\cal A}_{\rm short}$ at the $DD^*$ threshold to have the form
%-----------------
\begin{eqnarray}
{\cal A}_{\rm short}[B^+ \to (DD^*)^0_+ K^+] =
c_+ \, P \cdot \epsilon_{D^*} ,
\label{BtoDDK:short}
\end{eqnarray}
%-----------------
where $P$ is the 4-momentum of the $B^+$ and $\epsilon_{D^*}$
is the polarization 4-vector of the $D^*$. 
Heavy quark spin symmetry guarantees that the polarization vector 
$\epsilon_{D^*}$ can be identified with the polarization vector 
$\epsilon_X$ of the $X$.  The decay rate is obtained by
squaring the amplitude in Eq.~(\ref{BtoXK:fact}), 
summing over the spin of the $X$,
and integrating over phase space.  The resulting expression 
for the decay rate is
%-----------------
\begin{eqnarray}
\Gamma[B^+ \to  X K^+] =
| c_+ |^2 
{\lambda^{3/2}(m_B,m_X,m_K) \over 32 \pi m_B^3 m_X} \big| {\cal A}_X|^2,
\label{GamBtoXK}
\end{eqnarray}
%-----------------
where $\lambda(x,y,z)$ is the triangle function:
\begin{eqnarray}
\lambda(x,y,z)=x^4+y^4+z^4-2(x^2y^2+y^2z^2+z^2x^2).
\label{triangle}
\end{eqnarray}
The long-distance factor $|{\cal A}_X|^2$ is given in Eq.~(\ref{AXsq}).
The result in Eq.~(\ref{GamBtoXK}) was obtained in Ref.~\cite{Braaten:2004fk}
for the special case $\gamma_{\rm im} = 0$
and used to estimate the order of magnitude of the decay rate for 
$B^+ \to  X K^+$.  The estimate is consistent with the measurement
of the product of the branching fractions 
for $B^+ \to X K^+$ and $X \to J/\psi \, \pi^+ \pi^-$ \cite{Choi:2003ue}
provided $J/\psi \pi^+ \,\pi^-$ is one of the major decay modes of $X$.

The factorization formula for the decay rate for $B^0 \to  X K^0$
has the same form as in Eq.~(\ref{GamBtoXK}) except that the coefficient
$c_+$ in the short-distance decay amplitude in Eq.~(\ref{BtoDDK:short})
has a different value.  In Ref.~\cite{Braaten:2004ai}, it was pointed out 
that the decay rate for $B^0 \to  X K^0$ should be suppressed compared to 
$B^+ \to  X K^+$.  That suppression can be understood by considering the
short-distance amplitude for $B \to  D D^* K$.
The dominant contributions to most decay amplitudes of the $B$ meson  
are believed to be factorizable into the product of matrix elements 
of currents.  The factorizable contributions to the decay amplitude for
$B^+ \to  (DD^*)^0_+ K^+$ have three terms:  the product of 
$B^+ \to \bar D^{*0}$ and $\emptyset \to D^0 K^+$ matrix elements,
where $\emptyset$ is the QCD vacuum, 
the product of $B^+ \to \bar D^0$ and $\emptyset \to D^{*0} K^+$
matrix elements, and the product of $B^+ \to K^+$ and 
$\emptyset \to (DD^*)^0_+$ matrix elements.
The factorizable contributions to the decay amplitude for
$B^0 \to  (DD^*)^0_+ K^0$ have only one term:  
the product of $B^0 \to K^0$ and $\emptyset \to (DD^*)^0_+$ matrix elements.
Heavy quark symmetry implies that the $\emptyset \to (DD^*)^0_+$ 
matrix element vanishes at the $D^0 \bar D^{*0}$ threshold.
The decay $B^0 \to  (DD^*)^0_+ K^0$  near the $D^0 \bar D^{*0}$ threshold
must therefore proceed through nonfactorizable terms
in the decay amplitude.  The resulting suppression 
of the coefficient $c_+$ in the short-distance factor 
for $B^0 \to  (DD^*)^0_+ K^0$  
results in a suppression of the rate for $B^0 \to  X K^0$
relative to the rate for $B^+ \to  X K^+$.
In Ref.~\cite{Braaten:2004ai}, a quantitative analysis of Babar data 
on the branching fractions for $B \to D^{(*)} D^{(*)} K$ 
\cite{Aubert:2003jq} was used to estimate the suppression factor 
to be an order of magnitude or more.

\section{Production of $\bm{D^0 {\bar D^{*0}}}$ Near Threshold}
\label{sec:BDDstar}

It was pointed out in Ref.~\cite{Braaten:2004fk} that the identification 
of $X$ as a $D D^*$ molecule could be confirmed by observing a peak in the
invariant mass distribution for $D^0 \bar D^{*0}$ (or $D^{*0} \bar D^0$)
near the $D D^*$ threshold in the decay $B \to D D^* K$.  
The shape of that invariant mass
distribution was given for a real scattering length $a$.
The shape would be the same for any other short-distance production 
process.  In this section, we consider the effect of an imaginary part 
of the scattering length on the $D D^*$ invariant mass distribution
for a short-distance production process.  To be specific, we consider 
the short-distance production process $B \to D^0 \bar D^{*0} K$.

The separation between the long-distance scale $1/|a|$ and all the 
short-distance momentum scales in the decay $B \to D^0 \bar D^{*0} K$
can be exploited through a factorization formula for the T-matrix element:
%-----------------
\begin{eqnarray}
{\cal T}[B \to D^0 \bar D^{*0} K] =
{\cal A}_{\rm short}[B \to (DD^*)^0_+ K]
\times
{\cal A}(E) \, \sqrt{4 m_{D^0} m_{D^{*0}}} \, {1 \over \sqrt{2}} .
\label{BtoDDK:fact}
\end{eqnarray}
%-----------------
In the long-distance factor, ${\cal A}(E)$ is the universal amplitude 
in Eq.~(\ref{AE-complex}) and the factor $1/\sqrt{2}$ 
is the amplitude for $D^0 \bar D^{*0}$ to be in the channel 
$(D D^*)^0_+$ with the large scattering length.
The short-distance factor ${\cal A}_{\rm short}$ is the same as in the
factorization formula for $B \to X K$ in Eq.~(\ref{BtoXK:fact}).

%------------------------------------------------------------------------------------
\begin{figure}[t]
\includegraphics[width=8cm]{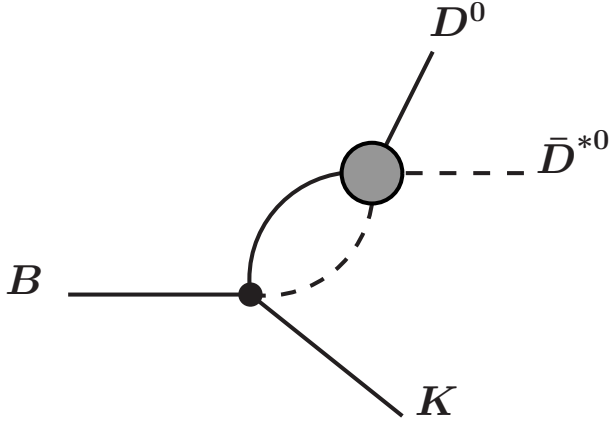}
\caption{
Feynman diagrams for $B \to D^{0} \bar D^{*0} K$ 
 that are enhanced near the $D^{0}\bar D^{*0}$
 threshold by a factor of $a m_\pi$. 
The blob represents the geometric series 
of diagrams shown in Fig.~\ref{fig:A-series}. 
\label{fig:BtoDDK}}
\end{figure}
%------------------------------------------------------------------------------------

The factorization formula in Eq.~(\ref{BtoDDK:fact})
can be motivated diagrammatically
by separating the loop integrals in the decay amplitude 
according to whether the  virtual particles are off their energy 
shells by less than or by more than $\Lambda^2/(2 \mu)$. 
There are terms in the T-matrix element for the decay $B \to D^0 \bar D^{*0} K$ 
that are enhanced near the $D D^*$ threshold by a factor of $a m_\pi$.  
These terms can be represented by the Feynman diagrams in 
Fig.~\ref{fig:BtoDDK} and can be expressed in the form
%-----------------
\begin{eqnarray}
{\cal T}[B \to D^0 \bar D^{*0} K] = 
{\cal A}^{(\Lambda)}[B \to (DD^*)^0_+ K] \, 
L(\Lambda,E) \, {\cal A}(E) \, \sqrt{4 m_{D^0} m_{D^{*0}}} \, {1 \over \sqrt{2}} .
\label{T:BtoDDK}
\end{eqnarray}
%-----------------
The first factor ${\cal A}^{(\Lambda)}$, which is represented 
by the dot in Fig.~\ref{fig:BtoDDK}, is an amplitude 
for the decay into $(DD^*)^0_+ K$ in which all virtual particles are
off their energy shells by more than $\Lambda^2/(2 \mu)$.
It is therefore insensitive to the relative momentum $\bm{p}$.
The second factor $L(\Lambda,E)$ is the amplitude for the propagation 
of the $D$ and $D^*$ between contact interactions, which is given  
in Eq.~(\ref{L-loop}).
The condition $|E| \ll \Lambda^2/(2 \mu)$ implies that  
the $\sqrt{-2 \mu E}$ term in $L(\Lambda,E)$ can be neglected.  
The resulting expression for the T-matrix element is the factorization 
formula in Eq.~(\ref{BtoDDK:fact}), with the short-distance factor 
${\cal A}_{\rm short}$ given in Eq.~(\ref{Ashort-BtoDDK}).

%------------------------------------------------------------------------------------
\begin{figure}[t]
\includegraphics[width=10.5cm]{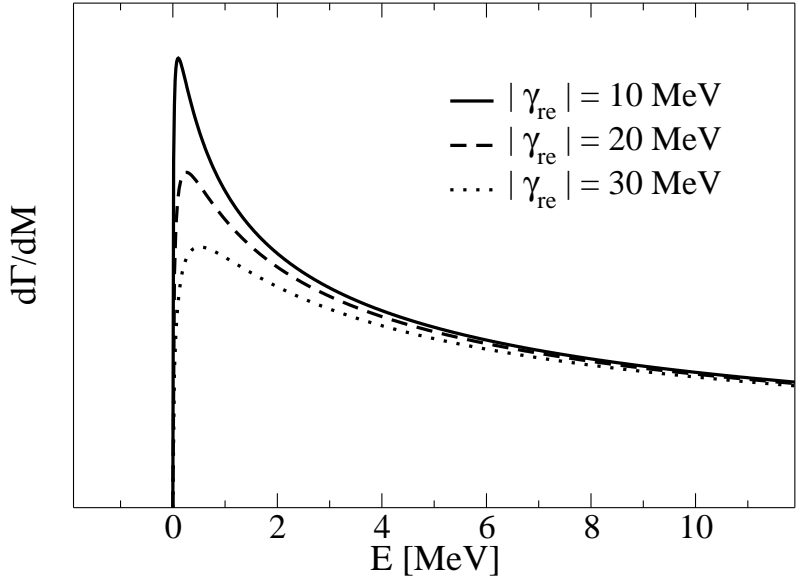}
\caption{
 The $D D^*$ invariant mass distribution in $B \to D^0 \bar D^{*0} K$ 
 for $\gamma_{\rm im} = 10$ MeV
 and various values of $|\gamma_{\rm re}|/\gamma_{\rm im}$.
 The horizontal axis is the difference $E=M-(m_{D^0}+m_{D^{*0}})$
 between the invariant mass $M$ and the $D^{0}\bar D^{*0}$ threshold. 
\label{fig:DDinvmass}}
\end{figure}
%------------------------------------------------------------------------------------

We proceed to use the factorized expression for the decay amplitude
in Eq.~(\ref{BtoDDK:fact}) to calculate the $D D^*$ invariant mass 
distribution near the $D^0 \bar D^{*0}$ threshold
in the decay $B^+ \to  D^0 \bar D^{*0} K^+$.
Lorentz invariance constrains the short-distance amplitude 
${\cal A}_{\rm short}$ at the $DD^*$ threshold to have the form
in Eq.~(\ref{BtoDDK:short}).  The decay rate is obtained by
squaring the amplitude in Eq.~(\ref{BtoDDK:fact}), 
summing over the spins of the $\bar D^{*0}$, and integrating over 
phase space.  The resulting expression for the differential decay rate 
with respect to the $D D^*$ invariant mass $M$ near the 
$D^0 \bar D^{*0}$ threshold is
%-----------------
\begin{eqnarray}
{d \Gamma \over d M} [B^+ \to  D^0 \bar D^{*0} K^+] =
| c_+ |^2 
{\mu \, \lambda^{3/2}(m_B,M,m_K) \over 256 \pi^3 m_B^3 M^2} \, 
\lambda^{1/2}(M, m_{D^0}, m_{D^{*0}})  \, |{\cal A}(E)|^2 ,
\label{dGam}
\end{eqnarray} 
%-----------------
where $E$ is the energy of the $D^0 \bar D^{*0}$ in its rest frame
relative to the $D^0 \bar D^{*0}$ threshold:
%-----------------
\begin{eqnarray}
E  = M - (m_{D^0} + m_{D^{*0}}) .
\label{EM-diff}
\end{eqnarray} 
%-----------------
We have used the fact that $M$ is near the $D^0 \bar D^{*0}$ threshold
to replace a factor of $m_{D^0} m_{D^{*0}}$ in Eq.~(\ref{dGam}) by
$\mu M$.
The result in Eq.~(\ref{dGam}) was obtained previously in 
Ref.~\cite{Braaten:2004fk} for the special case $\gamma_{\rm im}=0$.
For $M$ near the $D^0 \bar D^{*0}$ threshold, the only significant 
variation with $M$ is through the long-distance factor $|{\cal A}(E)|^2$ 
and the threshold factor
%-----------------
\begin{eqnarray}
\lambda^{1/2}(M, m_{D^0}, m_{D^{*0}}) \approx 2 M \sqrt{2 \mu E} .
\end{eqnarray}
%-----------------
If the complex scattering length is parameterized as in 
Eq.~(\ref{a-complex}), the long-distance factor is
%-----------------
\begin{eqnarray}
 |{\cal A}(E)|^2 = 
{ 4 \pi^2/\mu^2  \over 
( (2 \mu E)^{1/2} + \gamma_{\rm im} )^2 + \gamma_{\rm re}^2}.
\label{AE-sq:E>0}
\end{eqnarray} 
%-----------------

The shape of the $D^0 \bar D^{*0}$ invariant mass distribution 
in Eq.~(\ref{dGam}) is given by the factor 
$\sqrt{2 \mu E} \, |{\cal A}(E)|^2$.  Note that it depends on 
$\gamma_{\rm re}$ and $\gamma_{\rm im}$ but not on the sign of 
$\gamma_{\rm re}$.  The invariant mass distribution is shown 
in Fig.~\ref{fig:DDinvmass} for $\gamma_{\rm im} = 10$ MeV 
and three values of $|\gamma_{\rm re}|$: 10, 20, and 30 MeV.  
The peak in the invariant mass distribution
occurs at $E  = |\gamma|^2/(2 \mu)$, 
where $|\gamma| = \sqrt{\gamma_{\rm re}^2+\gamma_{\rm im}^2}$.
The value at the peak is proportional to 
$(|\gamma|+\gamma_{\rm im})^{-1}$.
The full width at half maximum is 
$2 (2|\gamma|+\gamma_{\rm im})
[(|\gamma|+\gamma_{\rm im}) (3|\gamma|+\gamma_{\rm im})]^{1/2}/\mu$.

\section{The $\bm{X}$ Line Shape}
\label{sec:Xlineshape}

The $X$ is observed as a peak in the invariant mass
distribution of its decay products, such as $J/\psi\,\pi^+\pi^-$.
Its mass and width are extracted 
from that invariant mass distribution.
For instance, the Belle collaboration obtained their value for the 
mass and the upper bound on the width by fitting the 
$J/\psi\,\pi^+\pi^-$ invariant mass distribution in 
$B^+ \to J/\psi\,\pi^+\pi^- \, K^+$ near the $D^0 \bar D^{*0}$ 
threshold to a resolution-broadened Breit-Wigner function 
on top of a polynomial background. 
The shape of the invariant mass distribution of the decay products 
of the $X$ is called the {\it line shape}.
The resonant interactions in the $D^{0}\bar D^{*0}/D^{*0}\bar D^{0}$
system can significantly modify the line shape, so it need not have the 
conventional Breit-Wigner form. 
In this section, we compute the line shape of the $X$ in short-distance 
decays of the $X$.  To be definite, we consider the production process
$B \to H K$, where $H$ is the hadronic system consisting of 
$J/\psi\,\pi^+\pi^-$ with invariant mass near the $D^0 \bar D^{*0}$ threshold.
However our results on the line shape will apply more generally to any
short-distance production process for $X$ and any short-distance
decay mode of $X$.

The separation between the long-distance scale $1/|a|$ and all the 
short-distance momentum scales in the decay $B \to H K$ can be 
exploited through a factorization formula for the T-matrix element:
%-----------------
\begin{eqnarray}
{\cal T}[B \to H K] 
= {\cal A}_{\rm short}[B \to (D D^*)^0_+ K] \times {\cal A}(E)
\times {\cal A}_{\rm short}[(D D^*)^0_+ \to H ] .
\label{BtoHK:fact}
\end{eqnarray} 
%-----------------
There is an implied sum over the spin states of the $D^*$.
The long-distance factor ${\cal A}(E)$ depends on the complex-valued 
scattering length $a$ and is given in Eq.~(\ref{AE-complex}).
Its argument $E$ is the difference between the invariant mass 
$M$ of the hadronic system $H$ and the $D^0 \bar D^{*0}$ threshold,
as given in Eq.~(\ref{EM-diff}).
The short-distance factor ${\cal A}_{\rm short}$ associated with the 
initial state is the same one that appears in the factorization 
formulas for $B \to X K$ in Eq.~(\ref{BtoXK:fact}) and for 
$B \to D^0 \bar D^{*0} K$ in Eq.~(\ref{BtoDDK:fact}).
The short-distance factor ${\cal A}_{\rm short}$ associated with the 
final state is the same one that appears in the factorization 
formulas for $X \to H$ in Eq.~(\ref{XtoH:fact}) and for 
$D^0 \bar D^{*0} \to H$ in Eq.~(\ref{DDtoH:fact}).

%------------------------------------------------------------------------------------
\begin{figure}[t]
\includegraphics[width=8cm]{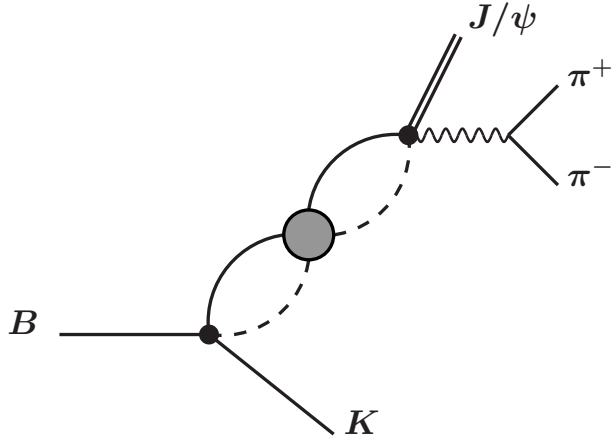}
\caption{
Feynman diagrams for $B \to J/\psi\,\pi^+\pi^-\, K$ 
 that are enhanced near the $D^{0}\bar D^{*0}$
 threshold by a factor of $a m_\pi$. 
The blob represents the geometric series 
of diagrams given in Fig.~\ref{fig:A-series}. 
\label{fig:Feyn1}}
\end{figure}
%------------------------------------------------------------------------------------

The factorization formula in Eq.~(\ref{BtoHK:fact})
can be motivated diagrammatically
by separating the loop integrals in the decay amplitude 
according to whether the virtual particles are off their energy 
shells by less than or by more than $\Lambda^2/(2 \mu)$. 
There are terms in the decay amplitude for $B \to H K$ 
that are enhanced by a factor of $a m_\pi$ when the invariant 
mass of $H$ is near the $D^0 \bar D^{*0}$ threshold.  
These terms can be represented by the Feynman diagrams in 
Fig.~\ref{fig:Feyn1} and can be expressed in the form
%-----------------
\begin{eqnarray}
 {\cal T}[B \to H K] 
&=& {\cal A}^{(\Lambda)}[B \to (D D^*)^0_+ K] \, L(\Lambda,E) \, 
{\cal A}(E) \, L(\Lambda, E) \,
{\cal A}^{(\Lambda)}[(D D^*)^0_+ \to H] .
\label{T:BtoHK}
\end{eqnarray} 
%-----------------
The factors ${\cal A}^{(\Lambda)}$, which are represented 
by dots in Fig.~\ref{fig:Feyn1}, are amplitudes 
in which all virtual particles are
off their energy shells by more than $\Lambda^2/(2 \mu)$.
The factors of $L(\Lambda,E)$, which is given in Eq.~(\ref{L-loop}), 
are the amplitudes for the propagation 
of the $D$ and $D^*$ between contact interactions.
The condition $|E| \ll \Lambda^2/(2 \mu)$ implies that  
the $\sqrt{-2 \mu E}$ term in $L(\Lambda,E)$ can be neglected.  
The resulting expression for the T-matrix element is the factorization 
formula in Eq.~(\ref{BtoHK:fact}), with the short-distance factors 
${\cal A}_{\rm short}$ given in Eqs.~(\ref{Ashort-BtoDDK}) 
and (\ref{Ashort-DDtoH}).

The factorization formula for the T-matrix element in (\ref{BtoHK:fact})
implies a factorization formula for the invariant mass distribution 
for the hadronic system $H$ near the $D^0\bar D^{*0}$ threshold.
If the hadronic system consists of particles with momenta $p_i$
and invariant mass $M$, the factorization formula is 
%-----------------
\begin{eqnarray}
&& \frac{d\Gamma}{dM}[B^+\to K^+ H]
\nonumber
\\
&=& 
{M \over 2 \pi m_B} 
 \int \big| {\cal A}_{\rm short}[B \to (D D^*)^0_+ K] \big|^2
(2 \pi)^4 \delta^4(P_B - P_H - P_K) 
{d^3 P_H \over (2 \pi)^3 2 E_H}{d^3 P_K \over (2 \pi)^3 2 E_K} 
\nonumber
\\
&& \times 
|{\cal A}(E)|^2 \times 
\int \big| {\cal A}_{\rm short}[(D D^*)^0_+ \to H ] \big|^2
(2 \pi)^4 \delta^4(P_H - \mbox{$\sum$}_i p_i) 
\prod_i {d^3 p_i \over (2 \pi)^3 2 E_i} .
\label{dGamdM:fact}
\end{eqnarray}
%-----------------
For $M$ near the $D^0 \bar D^{*0}$ threshold, the only significant 
variation with $M$ is through the long-distance factor 
$|{\cal A}(E)|^2$, where $E$ is the energy defined in Eq.~(\ref{EM-diff}).
If the complex scattering length is parameterized as in 
Eq.~(\ref{a-complex}), the long-distance factor is
%-----------------
\begin{subequations}
\begin{eqnarray}
 \bigl|{\cal A}(E)\bigr|^2 
&=& \frac{4\pi^2/\mu^2}{(|2\mu E|^{1/2}-\gamma_{\rm re})^2 + \gamma_{\rm im}^2} 
\hspace{2.1cm}       \;  E \le 0, 
\label{AEsq-pos}
\\
&=& \frac{4 \pi^2/\mu^2}{((2\mu E)^{1/2} + \gamma_{\rm im})^2 + \gamma_{\rm re}^2}
\hspace{2cm}          \; E\ge 0.
\end{eqnarray}
\label{AEsq}
\end{subequations}
%-----------------
This factor gives the line shape of the $X$.
Note that for $E>0$, the line shape does not depend on the sign of 
$\gamma_{\rm re}$.  However for $E<0$, the line shape is completely 
different for $\gamma_{\rm re}>0$ and $\gamma_{\rm re}<0$.

%------------------------------------------------------------------------------------
\begin{figure}[t]
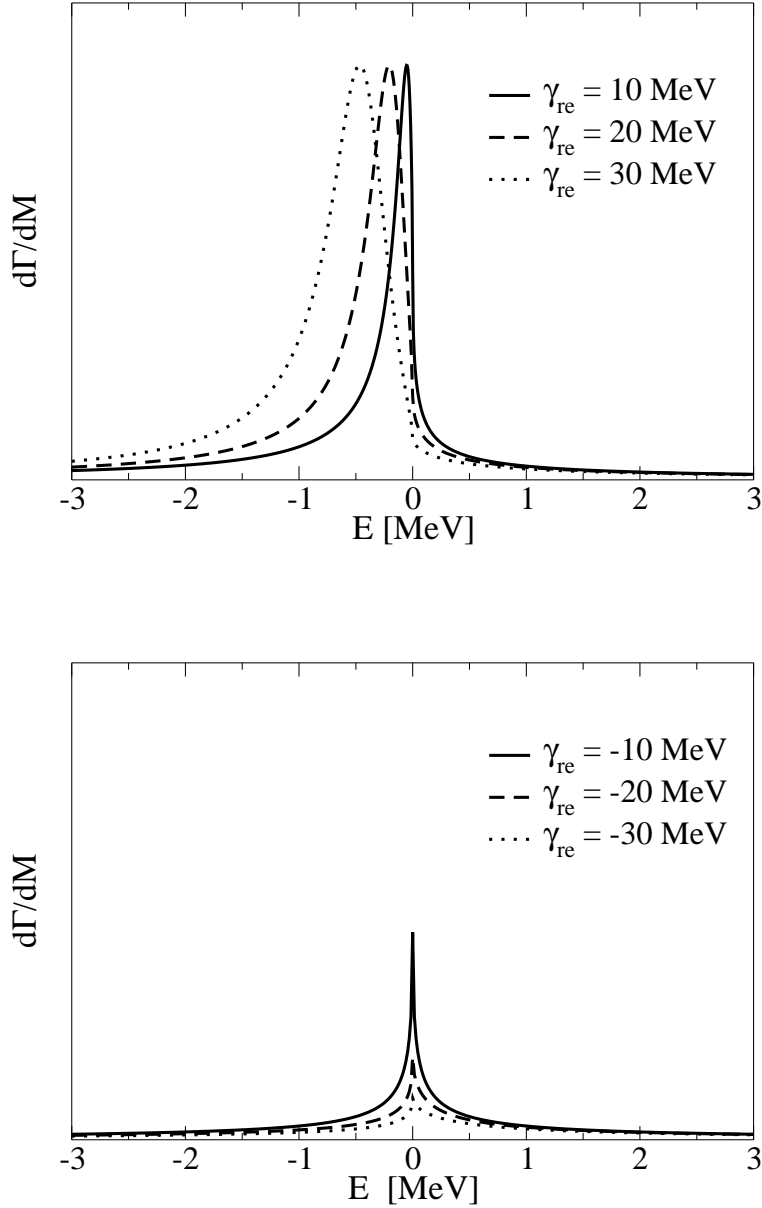

\includegraphics[width=10cm]{./Xls1.eps}

\vspace{1.5cm}

\includegraphics[width=10cm]{./Xls2.eps}
\caption{
 The $X$ line shape in a short-distance decay channel,
 such as $J/\psi \, \pi^+ \pi^-$, 
 for $\gamma_{\rm im} = 10$ MeV
 and for various positive values (upper panel) 
 and various negative values (lower panel) of 
 $\gamma_{\rm re}/\gamma_{\rm im}$.
 The horizontal axis is the difference $E=M-(m_{D^0}+m_{D^{*0}})$
 between the invariant mass $M$ and the $D^{0}\bar D^{*0}$ threshold. 
\label{fig:lineshape}}
\end{figure}
%------------------------------------------------------------------------------------

In the case $\gamma_{\rm re} > 0$,
the peak in the invariant mass distribution occurs below 
the $D^{0} \bar D^{*0}$ threshold  by the amount 
$\gamma_{\rm re}^2/(2 \mu)$.
The $X$ line shape is illustrated in the upper panel of 
Fig.~\ref{fig:lineshape} for $\gamma_{\rm im} = 10$ MeV 
and for three positive values of $\gamma_{\rm re}$: 10, 20, and 30 MeV.  
If $\gamma_{\rm im} < \gamma_{\rm re}$, the full width of the peak at 
half maximum is $2 \gamma_{\rm re} \gamma_{\rm im}/\mu$.
The line shape for $E<0$ is symmetric about the peak 
as a function of $|E|^{1/2}$ but not as a function of $E$.
If $\gamma_{\rm im} \ll \gamma_{\rm re}$,
the line shape in Eq.~(\ref{AEsq}) is sharply peaked at 
$E = - \gamma_{\rm re}^2/(2\mu)$ and it can be approximated 
by a delta function:
%-----------------
\begin{eqnarray}
|{\cal A}(E)|^2  \approx
{4 \pi^3 \gamma_{\rm re}  \over \mu^3 \gamma_{\rm im}}
\delta \big( E + \gamma_{\rm re}^2/(2 \mu) \big).
\end{eqnarray}
%-----------------
Note that the condition $\gamma_{\rm im} \ll \gamma_{\rm re}$
is equivalent to $\Gamma_X \ll E_X/4$.

In the case $\gamma_{\rm re}<0$, the peak in the invariant mass 
distribution occurs at the $D^{0} \bar D^{*0}$ threshold.
The $X$ line shape is illustrated in the lower panel of 
Fig.~\ref{fig:lineshape} for $\gamma_{\rm im} = 10$ MeV and for three 
negative values of $\gamma_{\rm re}$:  $-10$, $-20$, and $-30$ MeV.  
The line shape has a cusp at $E=0$.  Bugg has proposed that 
the $X$ can be identified with this cusp at the 
$D^{0} \bar D^{*0}$ threshold \cite{Bugg:2004rk}.
The normalization is the same in the upper and lower panels of 
Fig.~\ref{fig:lineshape}.  Note that the area under the cusp 
in the lower panel of Fig.~\ref{fig:lineshape} is much smaller 
than the area under the resonance in the upper panel
for the same values of $\gamma_{\rm im}$ and $|\gamma_{\rm re}|$.
Thus a cusp seems less likely as an interpretation for the $X(3872)$
than a resonance, although a quantitative analysis would be 
required to rule out that possibility.

The integral over all energies of the line shape of a 
conventional Breit-Wigner resonance is convergent.
In contrast, the integral of the line shape in Eq.~(\ref{AEsq})
diverges logarithmically as the endpoints $E_{\rm min}$
and $E_{\rm max}$ of the integral increase in magnitude. 
This follows from the fact that the line shape in Eq.~(\ref{AEsq})
decreases as $1/|E|$ for 
$(2 \mu |E|)^{1/2} \gg |\gamma_{\rm re}|,\gamma_{\rm im}$.
That expression for the line shape is of course only accurate 
for $|E|$ lower than
$\Lambda^2/(2 \mu) \sim 10$ MeV, where $\Lambda\sim m_\pi$ 
is the natural momentum scale for low-energy $D D^*$ scattering.  
Thus the logarithmic dependence on 
$E_{\rm min}$ and $E_{\rm max}$ holds only for
$|E_{\rm min}|,E_{\rm max} < \Lambda^2/(2 \mu)$.
It is convenient to define $p_{\rm min}$ and $p_{\rm max}$
by $E_{\rm min} = - p_{\rm min}^2/(2 \mu)$ and 
$E_{\rm max} = + p_{\rm max}^2/(2 \mu)$.  
The integral of the factor in Eq.~(\ref{AEsq}) reduces in the limit
$p_{\rm min}, p_{\rm max} \gg |\gamma_{\rm re}|, \gamma_{\rm im}$ to 
%-----------------
\begin{eqnarray}
\int_{E_{\rm min}}^{E_{\rm max}}  |{\cal A}(E)|^2 dE \approx
{4 \pi^2 \over \mu^3} \left( 
\log {p_{\rm min} p_{\rm max} \over \gamma_{\rm re}^2 + \gamma_{\rm im}^2}
+ {\pi \gamma_{\rm re} \over \gamma_{\rm im}} \, \theta(\gamma_{\rm re})
	- f(\gamma_{\rm re}/\gamma_{\rm im}) \right),
\end{eqnarray}
%-----------------
where $f(x)$ is the function
%-----------------
\begin{eqnarray}
f(x) = x \arctan(1/x) + (1/x)\arctan(x).
\end{eqnarray}
%-----------------
This function has a limited range, varying from 1 at $x=0$ 
and $x = \pm \infty$ to $\pi/2$ at $x = \pm 1$.

The factorization formula for the invariant mass 
distribution of $H$ in the case $\gamma_{\rm re} > 0$ has important 
implications for measurements of the branching fractions of $X$.
Since the two short-distance factors in Eq.~(\ref{dGamdM:fact})
are insensitive to $E$, we can set $M$ to $m_{D^0} + m_{D^{*0}}$ 
or to $m_X$ in those factors.
The short-distance factor associated with the decay of the $B^+$ 
reduces to $\Gamma[B \to X K]/(2 \pi |{\cal A}_X|^2)$.
If $\gamma_{\rm re} > 0$, the short-distance factor associated with the
formation of $H$ reduces to $\Gamma[X \to H]/|{\cal A}_X|^2$.
Thus the differential decay rate in 
Eq.~(\ref{dGamdM:fact}) reduces to
%-----------------
\begin{eqnarray}
{d \Gamma \over dM}[B \to H K]  =
\Gamma[B \to X K] \, {\rm Br}[X \to H] \;
{ \Gamma_X \, |{\cal A}(E)|^2 \over 2 \pi \, |{\cal A}_X|^4} .
\label{dGamdM:pos}
\end{eqnarray}
%-----------------
If the product of $\Gamma[B \to X K]$ and ${\rm Br}[X \to H]$
is measured by integrating $d\Gamma/dM$ over the energy interval 
from $E_{\rm min}$ to $E_{\rm max}$ with 
$|E_{\rm min}|,E_{\rm max} \gg E_X,\Gamma_X$, it 
will be in error by the factor
%-----------------
\begin{eqnarray}
{\Gamma_X \over 2 \pi |{\cal A}_X|^4}
\int_{E_{\rm min}}^{E_{\rm max}} |{\cal A}(E)|^2  dE
& \approx & 
\left[ 1 + {\Gamma_X^2 \over 16 E_X^2} \right]^{-1}
\nonumber
\\
&&  \hspace{-1cm}
\times \left[ 1 + 
\left( \log { |E_{\rm min}|^{1/2} E_{\rm max}^{1/2}
	\over E_X + \Gamma_X^2/(16 E_X)}
	-f(4E_X/\Gamma_X) \right) {\Gamma_X \over 4 \pi E_X} \right] .
\end{eqnarray}
%-----------------
The error would cancel in the ratio of the branching fractions 
for any two short-distance decay modes of $X$.  The error would not 
cancel in the ratio of the branching fractions 
for a short-distance decay mode of $X$ and one of the
long-distance decay modes $D^0 \bar D^0 \pi^0$ and 
$D^0 \bar D^0 \gamma$.  This effect should be taken into account 
in analyzing the decays of the $X(3872)$.

\section {Summary}
\label{sec:summary}

If the $X(3872)$ is a loosely-bound S-wave molecule corresponding to a
$C=+$ superposition of $D^0\bar D^{*0}$ and $D^{*0}\bar D^{0}$, 
the scattering length $a$ in the $(D D^*)_+^0$
channel is large compared to all other length scales of QCD.
The decays of the $X$ implies that the large scattering length
has an imaginary part.  It can be conveniently parameterized 
in terms of the real and imaginary parts of $1/a$ as in 
Eq.~(\ref{a-complex}).  The binding energy $E_X$ and the width $\Gamma_X$ 
are expressed in terms of those parameters in 
Eqs.~(\ref{EX-complex}) and (\ref{Gam-uni}).

The large scattering length can be exploited through factorization 
formulas for decay rates of $X$.
For short-distance decay modes that do not proceed through the decay of a
constituent $D^*$ of the $X$,
the long-distance factor in the factorization formula 
is proportional to $1/|a|$ and is given in Eq.~(\ref{AXsq-EGam}).  
If a partial width of the $X$
is calculated using some model with a specific binding energy 
for the $X$, the factorization formulas can be used to extrapolate 
the prediction to other values of the binding energy
and to take into account the width of the $X$.

The large scattering length can also be exploited through factorization 
formulas for production rates of $X$, 
$D^0\bar D^{*0}$ near threshold, $D^{*0}\bar D^{0}$ near threshold, 
and decay products of $X$ with invariant mass near the 
$D^0\bar D^{*0}$ threshold.
The long-distance factor in the factorization formula for 
production rates of $X$ is proportional to $1/|a|$.
For production of $D^0\bar D^{*0}$ and $D^{*0}\bar D^{0}$ near threshold,
the factorization formula implies that the dependence on the invariant 
mass is through the factor $\sqrt{2 \mu E}|{\cal A}(E)|^2$,
where $E$ is the invariant mass relative to the $D^0\bar D^{*0}$
threshold and $|{\cal A}(E)|^2$ is given in Eq.~(\ref{AE-sq:E>0}).
The peak in the invariant mass distribution is above the threshold 
by the amount $E_X + \Gamma_X^2/(16 E_X)$.
The line shape of the $X$ can be measured through the 
invariant mass distribution of its decay products.
In the case of short-distance decay modes, the factorization formulas 
imply that near the $D^0\bar D^{*0}$ threshold, the shape of the 
invariant mass distribution is given by the factor $|{\cal A}(E)|^2$
in Eq.~(\ref{AEsq}).  If Re$(a)<0$, the distribution has a cusp at $E=0$,
as shown in the lower panel of Fig.~\ref{fig:lineshape}.
If Re$(a)>0$, the distribution has a peak at $E = - E_X$,
as shown in the upper panel of Fig.~\ref{fig:lineshape}.
In contrast to a Breit-Wigner resonance, the integral over the line shape is
logarithmically sensitive to the endpoints of the integration region.
This effect should be taken into account in analyzing the
production and decay of the $X$. 

\begin{acknowledgments}
% put your acknowledgments here.
This research was supported in part by the Department of Energy 
under grant DE-FG02-91-ER4069.  
\end{acknowledgments}

%%%%%%%%%%%%%%%%%%%%%%%%%%%%%%%%%%%%%%%%%%%%%%%%%%%%%%%%%%%%%%%%%%%%%%%%%%%%
% Create the reference section using BibTeX:
%----------------------------------------------------------------------

\end{document}